\documentclass[final,5p]{elsarticle}

\usepackage{natbib,geometry,graphicx,hyperref}
\usepackage{amssymb,amsmath,xspace,tabularx,pifont,floatrow}
\usepackage{multicol} 
\usepackage{stfloats}

\def\pb{$^{210}$Pb\xspace}
\journal{Nuclear Instruments and Methods}

\newcommand{\Pb}{$^{210}$Pb }

\newcommand{\Po}{$^{210}$Po }
\newcommand{\Rn}{$^{222}$Rn }

\biboptions{numbers,sort&compress}

\begin{document}

\begin{frontmatter}

\title{Evaluation and mitigation of trace \pb contamination on copper surfaces}
\author[PNNL]{R.~Bunker}
\author[SLAC]{T.~Aramaki}
\author[PNNL]{I.J.~Arnquist}
\author[SMU]{R.~Calkins\corref{rec}}\ead{rcalkins@smu.edu}
\author[SMU]{J.~Cooley}
\author[PNNL]{E.W.~Hoppe}
\author[PNNL]{J.L.~Orrell}
\author[PNNL]{K.S.~Thommasson}

\address[PNNL]{Pacific Northwest National Laboratory, Richland, WA 99352, USA}
\address[SLAC]{SLAC National Accelerator Laboratory/Kavli Institute for Particle Astrophysics and Cosmology, Menlo Park, California 94025, USA}
\address[SMU]{Department of Physics, Southern Methodist University, Dallas, Texas 75275, USA}
\cortext[rec]{Corresponding author}

\begin{abstract}
Clean materials are required to construct and operate many low-background physics experiments. 
High-purity copper has found broad use because of its physical properties and availability. 
In this paper, we describe methods to assay and mitigate \pb contamination on copper surfaces, such as from exposure to environmental radon or coming from bulk impurities. 
We evaluated the efficacy of wet etching on commercial samples and observed that \Po contamination from the copper bulk does not readily pass into solution. During the etch, the polonium appears to trap at the copper-etchant boundary, such that it is effectively concentrated at the copper surface. 
  We observed a different behavior for \pb;  
high-sensitivity measurements of the alpha emissivity versus time indicate the lowest level of \pb contamination ever reported for a commercial copper surface:  $0\pm12$\,nBq/cm$^2$ (1$\sigma$).  Additionally, we have demonstrated the effectiveness of mitigating trace \pb \textit{and} \Po surface backgrounds using custom, high-purity electroplating techniques. 
These approaches were evaluated utilizing assays performed with an XIA UltraLo-1800 alpha spectrometer.

\end{abstract}


\end{frontmatter}

\section{Introduction}\label{sec:intro}

Several new rare-event searches with improved sensitivity to interactions from dark matter and neutrinos are currently in their planning or construction phases (see, e.g., Refs.~\cite{Aalseth:2017fik,Kharusi:2018eqi,sensitivity}).
For these experiments to be successful, radioactive backgrounds must be controlled to exceptionally low levels. Standard shielding techniques can provide sufficient protection from most environmental backgrounds during experiment operations. As a result, radiocontaminants in the detector materials tend to be the dominant sources of expected background (see, e.g., Refs.~\cite{sensitivity,Orrell:2017rid}). To maximize sensitivity to new physics, advanced techniques need to be developed to not only improve the radiopurity of materials but to also validate them for use in experiments. 

In this article, we turn our attention to copper.  Copper is a commonly used material because it has a useful combination of physical properties and because high-purity copper is inexpensive and readily available from commercial sources. For example, copper is used as a detector packaging material in the CUORE~\cite{Alessandria:2012zp}, DAMIC~\cite{Aguilar-Arevalo:2016ndq}, M\textsc{ajorana}~\cite{Abgrall:2016cct}, and SuperCDMS~\cite{sensitivity} experiments.  Copper is also used in the designs of the DarkSide-20k~\cite{Aalseth:2017fik}, NEWS-G~\cite{Arnaud:2017bjh} and nEXO~\cite{Kharusi:2018eqi} detectors.

A specific background concern associated with copper is surface contamination.  Radiocontaminants can accumulate on copper surfaces when they come into contact with laboratory air containing radon or with other materials. For example, exposure to the average radon concentration  in the SNOLAB underground laboratory~\cite{SNOLAB_handbook} 
 can result in significant levels of surface contamination~\cite{Stein:2017tel}.
The immediate progeny of \Rn can plate out onto surfaces where they undergo a series of relatively prompt alpha and beta decays, leading to the long-lived radioisotope \Pb (22.2~year half-life~\cite{SHAMSUZZOHABASUNIA2014561}). Products of the \Pb decay chain include recoiling nuclei, alpha particles, x-rays, and beta particles, which can also produce Bremsstrahlung photons and neutrons from ($\alpha$,$n$) reactions. When \Pb and its progeny decay on or near an active detector medium, they can mimic a rare event signal. There is an additional concern from materials directly surrounding or in contact with the detector medium (as copper often is); contamination on `line-of-sight' surfaces has no shielding between the decay and the detector such that decay products can travel unimpeded into the active medium. In order to meet sensitivity goals, many current and future experiments need to specifically address and characterize this source of background for their copper components.

The decay sequence from \Rn to \Pb includes alpha decays.  Recoiling progeny from alpha decays occurring on copper surfaces have sufficient energy to be driven (`implanted') into the copper. Consequently, \Pb surface contamination due to exposure to radon can be implanted to depths of tens of nanometers~\cite{ZUZEL2012140}.  Due to this depth and the long half-life, implanted \Pb is a particularly challenging background. 
In general, conventional surface cleaning methods do not provide adequate mitigation (e.g., rinsing, wiping or sonicating with a mild solvent or alkaline detergent). Further, many next-generation experiments have such strict background requirements that it is challenging to measure (and thus validate) \Pb contamination levels with sufficient sensitivity. In this article, we describe our approaches to reducing \Pb surface backgrounds through cleaning and electroplating of copper surfaces, including use of a high-sensitivity assay method for evaluating the performance of our methods in the relevant low-level regime.

We describe preparation of our copper samples and the surface treatments in Section~\ref{sec:cu}.
Section~\ref{sec:xia} details our high-sensitivity measurement technique using an XIA UltraLo-1800 alpha spectrometer, including measurement protocols, characterization of the instrumental background, and assay results for our copper samples. Our attempt to also measure the \pb activity in the bulk of our copper samples is reported in Section~\ref{sec:bulk210Pb}. The paper concludes with a discussion of the results.

\section{Copper samples and surface treatments}\label{sec:cu}
The copper samples and surface treatments described in this section represent the development and validation of a program for post-fabrication cleaning of low-background copper components for rare-event searches.

\subsection{Commercially sourced copper}\label{sec:cu:vendor}
Based on the requirements of the SuperCDMS SNOLAB dark matter experiment~\cite{sensitivity}, we selected oxygen-free high-thermal conductivity (OFHC) copper for our commercially sourced copper samples. OFHC copper has good cryogenic properties and has been measured to have low radioisotope concentrations in the bulk material (see, e.g., Ref.~\cite{APRILE201143}). In order to maximize the sensitivity of the measurements presented in the next section, we fabricated samples with relatively large surface areas of $\mathcal{O}($1000\,cm$^2)$, each constructed from a set of smaller square plates to facilitate handling during the surface treatments. OFHC copper from two different vendors was used to fabricate two sets of square plates.

The first copper sample was fabricated from OFHC copper (alloy 101) purchased from McMaster-Carr, a U.S.-based distributor. No attempt was made to trace back to the originating manufacturer. Henceforth we will refer to this sample as the `McMaster copper.'
Four 6- by 6- by 0.25-inch plates were purchased, and new tooling and fresh machining fluids were used to mill $\sim$1/16 of an inch from both sides of each plate. Additionally, the edges of each plate were trimmed.  Directly following machining, an initial cleaning was performed to remove machining fluids, particulates (`dirt'), and copper oxide; each plate was degreased in a vapor degreaser for 20 minutes, immersed in an alkaline detergent for 5 minutes, and immersed in 50\% hydrochloric acid for one minute, with deionized-water rinses after each step.  
Following this initial cleaning, the plates were placed in clean zip-lock bags and transported to a class-10 cleanroom, where the cleaning procedure described in Sec.\,\ref{sec:cu:clean} was performed. 

A second sample was fabricated from OFHC copper obtained from a specific manufacturer, Aurubis (Hamburg, Germany). Based on the consistently low concentrations of U and Th measured in the bulk material (see, e.g., Refs.~\cite{Abgrall:2016cct,APRILE201143,Leonard:2007uv}), Aurubis was selected as the preferred copper manufacturer for construction of the SuperCDMS SNOLAB apparatus~\cite{sensitivity}. 
For this second sample, 4$\times$4 square-inch plates were cut from 6"-diameter rod stock---a more relevant form factor for fabrication of detector components (e.g., SuperCDMS detector housings). A total of nine plates were fabricated using the same machining and initial-cleaning protocols as for the McMaster copper; thus achieving a total sample surface area of 12$\times$12 square inches (as with the McMaster plates), but starting with a smaller initial plate thickness of $\sim$0.2\,inches.

\subsection{Wet etching}\label{sec:cu:clean}

A surface treatment was applied to each of the copper samples in order to remove surface radiocontaminants---in particular, any trace levels of \pb resulting from exposure to environmental radon during the fabrication and initial cleaning. To ensure removal of \pb implanted up to tens of nanometers, we chose an acidified-peroxide wet etching recipe developed at Pacific Northwest National Laboratory (PNNL) specifically for copper~\cite{HOPPE2007486,HOPPE2008}. Compared to a nitric-acid etch, the acidified peroxide allows for a more consistent and well-controlled removal of material.  While the test results in Ref.~\cite{HOPPE2007486} show comparable performance (relative to nitric acid) for removal of surface $^{209}$Po, they are not necessarily indicative of the mitigation performance for implanted \pb in the low-level regime that is of interest to the low-background physics community. Thus, our goal was to demonstrate that the PNNL recipe can be used to prepare a commercial copper surface with  $\mathcal{O}($nBq/cm$^2)$ levels of \pb.\footnote{For reference, the \pb surface contamination rate due to exposure to a typical level of environmental radon---$\mathcal{O}($10\,Bq/m$^3)$---and assuming an effective plate-out height of 10\,cm is approximately 10\,nBq/cm$^2$ per day of exposure.}

Each copper plate 
was placed in an acidified-peroxide bath consisting of a 1:3:96 mixture of Optima\texttrademark grade $\text{H}_{\text{2}}\text{SO}_{\text{4}}$, ACS grade Sn-stabilized $\text{H}_{\text{2}}\text{O}_{\text{2}}$, and $>$18\,M$\Omega$ pure water, respectively. 
The plates were continuously agitated while in the bath until oxides were removed and for not less than one minute, corresponding to removal of 1--2 microns of copper thickness from all surfaces. Each plate was then rinsed in pure water before being placed in a 1\% citric acid solution and agitated for approximately one minute. The citric acid passivates the surface of the copper to prevent formation of oxides. The plates were again rinsed in pure water and immediately blown dry with a stream of dry nitrogen. Once dry, they were placed in a nitrogen-purged vacuum oven set at 80$^{\circ}$C  and baked overnight to drive off any residual moisture and make the passivated coat more resilient. Following this treatment, each set of plates was shipped to Southern Methodist University (SMU) where the measurements described in Sec.\,\ref{sec:xia} were performed.  To protect the plates from exposure to radon and dust, they were heat sealed inside two layers of nitrogen-purged, class-10 nylon bagging prior to shipment.  Between measurements, the plates were stored in the heat sealed bags and additionally placed in a nitrogen-purged cabinet for redundant protection from radon.

\subsection{Use of electroformed copper}\label{sec:cu:ef}
In addition to the OFHC plates, copper sheets were electroformed using the same PNNL ultra-low-background method~\cite{HOPPE2014116} used to grow highly radiopure copper for the M\textsc{ajorana} experimental program~\cite{Abgrall:2016cct}. For the work described here, large-area copper sheets with a thickness of 200--300\,microns were electroformed in the PNNL shallow underground laboratory~\cite{ISI:000312034400015}.  

These electroformed sheets were used primarily to characterize the background performance of our UltraLo-1800 spectrometer. Prior measurements of PNNL electroformed copper 
indicate exceptionally low concentrations of U and Th in the bulk material (via high-sensitivity mass spectrometry)~\cite{ISI:000348040900014}, as well as surface alpha emissivities close to or consistent with expected sensitivity limits of other UltraLo-1800 spectrometers~\cite{private_xia,private_ibm}. 
In Ref.~\cite{ABE2018157}, \pb and $^{210}$Po were undetectable in the bulk material of commercially grown electroformed copper, whereas relatively large \pb and $^{210}$Po concentrations were measured in OFHC copper samples. 
Therefore, we expect little to no alpha activity from the electroformed copper, making the electroformed sheets nearly ideal for characterizing the UltraLo-1800's irreducible (`instrumental') background. The electroformed sheets also act as a control for evaluating the performance of the acidified-peroxide etch, because each sheet was subjected to the same cleaning and packaging protocols as the OFHC copper plates (cf.~Sec.\,\ref{sec:cu:clean}).

Two distinct electroformed-sheet geometries were produced, corresponding approximately to the shapes and sizes of the counting anodes in the UltraLo-1800:  a disk shape with a diameter of $\sim$30\,cm, and a square shape with an area of $\sim$45$\times$45\,cm$^2$. In both cases, the electroformed copper sheet was sized to be somewhat larger than the corresponding UltraLo-1800 electrode.  Thus, when centered on the instrument's sample tray, the disk (square) shaped sheet effectively covers the part of the sample tray that can contribute background events when reading out the instrument's corresponding inner (full) electrode. Ultimately, these electroformed sheets serve two purposes.  First, covering the sample tray with electroformed copper enables characterization of the instrument's background without contributions from the surface of the sample tray, because the electroformed copper shields any alphas emitted from the tray while emitting few (to none) of its own.  Second, these electroformed sheets can serve as low-background tray liners for sample assays requiring the instrument's best possible sensitivity.\footnote{Note that this second functionality was not utilized for the sample results reported here, because all of our copper samples were fabricated to have large enough areas to shield the spectrometer's readout anode from any alphas emitted from the sample tray.}

\subsection{Electroplating}\label{sec:cu:plate}
We also explored use of the PNNL electroforming technique as a method for mitigating radiocontaminants on the surfaces of copper parts.  The idea is to electroplate a thin layer of highly radiopure copper onto the surfaces of copper parts fabricated from less-radiopure OFHC copper. The ultra-high-purity electroplated layer then acts to shield radiation emitted by radiocontaminants on the OFHC copper surfaces.  This is expected to be particularly effective for shielding detectors from short-range radioactivity, such as the decay products from the \pb decay chain (e.g., alphas, $^{206}$Pb ions, and low-energy betas and x-rays). For cases in which fabricating detector components from the higher-purity PNNL electroformed copper is not practical (e.g., due to geometry or requisite thickness), this new electroplating treatment represents a cost-effective alternative for mitigating backgrounds associated with OFHC copper.

Our OFHC copper plates offered an opportunity to test this new electroplating surface treatment. As will be discussed in the next section, following wet etching with the acidified peroxide, we measured a significant rate of $^{210}$Po surface activity.  
As a result, the McMaster plates were subjected to the electroplating treatment and subsequently remeasured two more times; results are discussed in Sec.\,\ref{ssec:mcmaster}.
It is worth noting that each of the four McMaster plates was first re-etched with the acidified peroxide prior to electroplating, which likely `reset' the $^{210}$Po surface activity back to (approximately) the initial level measured after the first wet etch.  The likely source of this $^{210}$Po contamination is from the bulk of the OFHC copper. During the etch, it appears that \Po from the bulk is trapped at the copper-etchant boundary, such that it is effectively concentrated at the copper surface. 
Each plate was submerged in an electroplating bath where it was first electropolished---to prepare the surface for copper growth---and then $\sim$0.5\,mm of copper was electroplated using the PNNL technique. A relatively thick layer was electroplated to demonstrate process uniformity.  This surface treatment can therefore be tailored to the thickness most applicable to shielding against radiocontaminants of concern.
To ensure that the plates would not be too thick for the UltraLo-1800 spectrometer's sample-tray opening, they were subsequently milled down to an overall thickness of $\sim$4.5\,mm, yielding an outer layer of ultra-high-purity electroplated copper at least 100\,microns thick.
Finally, each plate was again wet etched and packaged following the full set of procedures outlined in Sec.\,\ref{sec:cu:clean}.

\begin{figure}
\centering
\includegraphics[width=0.95\textwidth]{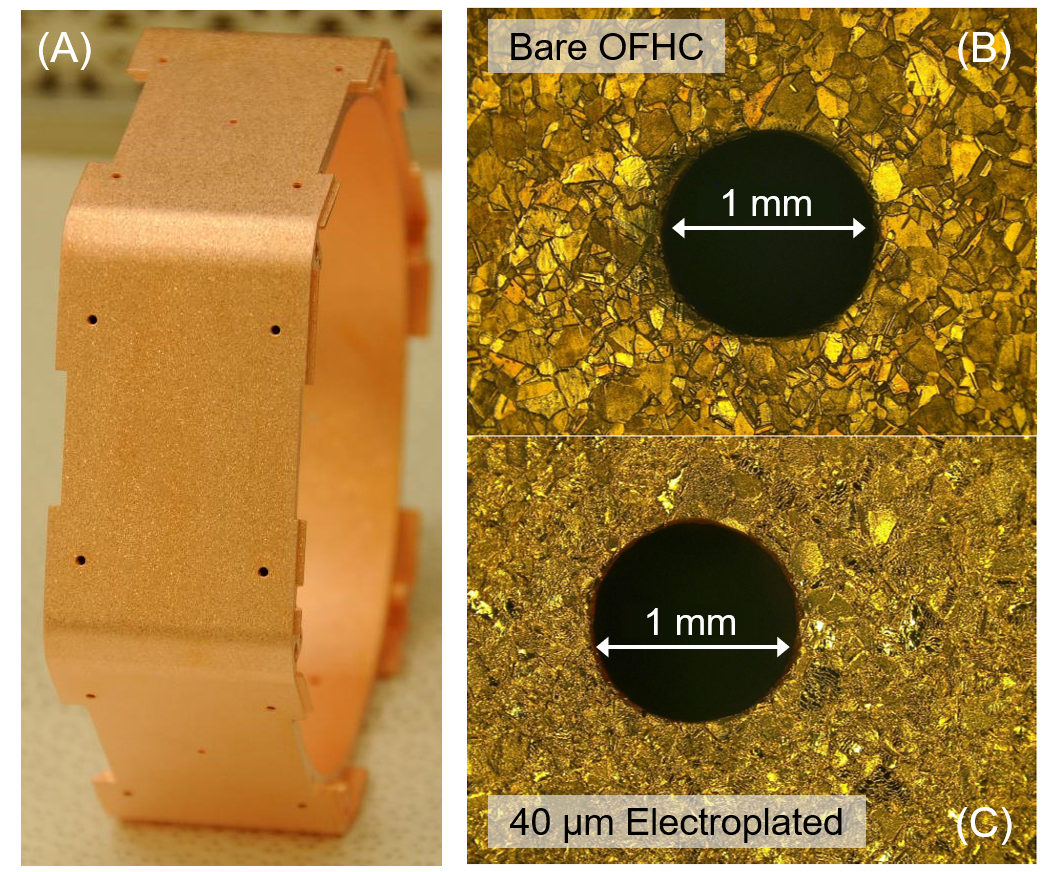}
\caption{(A)~Photo of a SuperCDMS SNOLAB detector housing after application of the PNNL electroplating surface treatment. (B)~Magnified image of one the screw holes prior to the electroplating treatment. The relatively large crystal grains characteristic of OFHC copper are apparent. (C)~Magnified image of the same screw hole after electroplating a 40\,micron layer of copper using the PNNL method.  The large OFHC grains have clearly been covered by the electroplated copper which grows in much smaller grains.}
\label{fig:housing}
\end{figure}

Electroplating to flat square plates is relatively straightforward.  To demonstrate that this method can be effective for detector components with a realistic geometry, we applied it to a SuperCDMS SNOLAB detector housing~\cite{sensitivity}. The SuperCDMS housings are fabricated from OFHC copper, have a hexagonal shape, and have several fine details including small screw holes.  Specifically, we sought to test if the electroplated copper would grow evenly and uniformly over the OFHC copper surface despite the more complex electric field created in the electroplating bath due to the nontrivial geometry of the detector housing. Figure~\ref{fig:housing} shows the results.  When the surface is viewed under magnification, it is clear that the OFHC copper surface has been fully covered by the electroplated copper, because the large grain sizes characteristic of OFHC copper are no longer visible underneath the electroplated copper's much smaller grain sizes.  Magnified images were taken of several locations, including (in particular) in and around the small screw holes.  In each case, the images show good coverage (as in Fig.~\ref{fig:housing}) of the OFHC copper surface for an estimated thicknesses of electroplated copper of $\sim$40\,microns, which is based on the integrated plating current and the approximate surface area of the housing.

\section{High-sensitivity alpha spectrometry}\label{sec:xia}

SMU operates an XIA UltraLo-1800 alpha spectrometer~\cite{xia} in a cleanroom located in a surface laboratory on the SMU campus. The UltraLo-1800 is an ionization chamber that can accommodate large-area samples up to 42$\times$42\,cm$^{2}$. 
The top of the chamber (opposite the sample tray) is instrumented with two configurable counting anodes: a square `full tray' anode with an area of 1800\,cm$^2$ and a smaller inner circular `wafer' anode with an area of 707\,cm$^2$.
The perimeter of these counting anodes is instrumented with a veto anode referred to as the `guard' channel. 
A potential of 1100\,V is applied between the grounded sample tray and the anodes. 
The volume of the chamber is filled with boil-off gas supplied by a liquid-argon dewar.  Alpha particles emitted from the surface of a sample ionize the argon count gas, and as the charges drift upward (due to the potential) they induce currents in the anodes. 
The UltraLo-1800 fits the induced pulse and uses pulse-shape discrimination (PSD) to characterize the energy of the alpha particle and the ‘rise time’ of the pulse. The risetime is indicative of the height above the sample tray from which the alpha originated.

When samples are changed, the counting chamber is exposed to the laboratory air which introduces moisture and environmental radon. Each measurement (or `run') begins with a purge during which the flow rate of argon into the chamber is increased to flush out contaminants. 
Additionally, we reject the first six hours of data after a purge, because in our experience additional time is required for the count gas to stabilize.

\subsection{Spectrometer backgrounds}

For materials with low surface activities, such as those used in dark matter and neutrinoless double-beta decay experiments (and under study here), the number of observed decays is low even for long counting times. Consequently, instrumental backgrounds are non-negligible. 
The primary backgrounds are from alpha decays that originate from the count gas or interior surfaces of the counting chamber. 
Because the SMU UltraLo-1800 is located in an above-ground surface laboratory, there is also a contribution from cosmic rays. 

The key variable for discriminating a sample's surface activity from alpha-emitting backgrounds in the count gas is the rise time of the anode pulses.
Decays occurring \textit{in} the count gas have a shorter rise time because the charge has less distance to drift until it is absorbed by the anode. 
We reject such `mid-air' alphas by excluding events with a rise time less than 60\,$\mu$s, and we estimate the residual background rate by measuring the high-purity electroformed sheets described in Sec.\,\ref{sec:cu:ef}; background measurements are discussed below in Sec.\,\ref{sec:PNNL_EF}.

Alphas originating from the chamber's interior surfaces can be discriminated in some cases. 
Alphas emitted from the sidewalls of the chamber ionize the gas under the guard anode, which is used to veto these events. 
Alphas originating from the anodes themselves have a very short drift distance and are thus rejected by the rise-time selection.

If a sample's surface area is smaller than the counting anode, a portion of the sample tray is exposed directly below the anode. 
Any alphas emitted from the surface of the sample tray represent a background that cannot be rejected because their locations (relative to the anodes) and rise times are comparable to alphas emitted from the sample. 
We can estimate the rate of these events by performing runs without a sample.  In general, we cover the sample tray with a conductive Teflon liner, which has been demonstrated to have a lower background rate than the bare stainless-steel surface of the sample tray. 

In this paper, our focus is \Pb and its progeny. In particular, we use decays of \Po as our primary measure of surface activity. A sample's \Pb surface activity can be inferred from several measurements of its \Po alpha emissivity separated in time.  We further enhance the sensitivity of our measurements by defining an alpha energy region of interest (ROI), thus effectively excluding instrumental backgrounds that give events with inconsistent energies. The \Po alpha energy is 5.3\,MeV~\cite{KONDEV20081527} and the quoted energy resolution of the UltraLo-1800 is $\sim$200\,keV (at 5\,MeV)~\cite{xia}, which we have confirmed with a 1.4\,Bq $^{230}$Th calibration source.  
We define a 1 MeV wide ROI centered around the 5.3\,MeV \Po alpha peak, which is the ROI referred to hereafter.

\subsection{Background modeling and limit setting}
\label{ssec:subtraction}
We estimate the spectrometer's backgrounds from dedicated measurements. 
We consider the background originating from the sample tray separately from non-tray backgrounds, because the tray component depends on the sample size. A separate model for the tray background allows us to scale by the fraction of the tray's surface that is exposed below the counting anode. 
We refer to the non-tray background as the `instrumental' background, because it is independent of the sample area. 
We utilize these two background models in a likelihood-based approach to obtain a best estimate for a sample's surface alpha activity.

\begin{table*}
\begin{tabular}{|c|c|c|}
\hline
Parameter & Calculation & Description \\
\hline
$\beta_1$ & $(1-\frac{\mathrm{sample\,area}}{\mathrm{counting\,area}})\frac{\mathrm{sample\,run\,time}}{\mathrm{tray\,run\,time}}$ & Scaling between sample run and tray-background measurement\\
$\beta_2$ & $\frac{\text{tray run time}}{\text{instrumental-background  run time}}$ & Ratio of tray- to instrumental-background measurement times \\
$\beta_3$ & $\frac{\text{sample run time}}{\text{instrumental-background  run time}}$ & Ratio of sample to instrumental-background measurement times \\
\hline
\end{tabular}
\caption{Dimensionless parameters $\beta_i$ used to scale contributions from the tray and instrumental backgrounds in the likelihood in Eq.~\ref{eq:like}.}
\label{tab:tau}
\end{table*}

The lengths of the background measurements and sample runs are different.  To construct our likelihood, we therefore define scaling parameters, $\beta_i$, which are summarized in Table~\ref{tab:tau}. 
We follow the general procedure outlined in Ref.~\cite{Rolke:2004mj} to define a likelihood function that incorporates the uncertainty from the background measurements.  Contributions from the sample and from the backgrounds all follow Poisson probabilities:
\begin{equation}
\label{eq:poisson}
\mathrm{Pois}( x | \lambda ) = e^{-\lambda} \left( \lambda x \right) / x!,
\end{equation}
where $\lambda$ is the expected mean and $x$ is the observed value in a single experiment. Our likelihood function is therefore a product of three probabilities:
\begin{equation}
\begin{aligned}
\label{eq:like}
{\cal L} =  & \mathrm{Pois}(N_\mathrm{obs} | s + \beta_1 b_{T} + \beta_3 b_{I}) \\
 & \times \mathrm{Pois}(N_{\mathrm{instr}}|b_{I}) \times \mathrm{Pois}(N_{\mathrm{tray}} | b_{T} + \beta_2 b_I ).
\end{aligned}
\end{equation}
$N_\mathrm{instr}$ is the measured number of instrumental-background events and $b_I$ is the estimate of the true number.  $N_\mathrm{tray}$ is the number of events observed in the tray-background measurement, with a corresponding expected mean that includes both the true number of tray-background events $b_T$ and a scaled contribution from the instrumental background. Finally, $N_\mathrm{obs}$ is the number of events observed in the sample run, with an expected mean equal to the sum of the true number of sample events $s$ and scaled contributions from both the tray and instrumental backgrounds.
We assume Poisson uncertainties for this likelihood, and it is trivial to 
maximize by setting $b_I = N_{\mathrm{instr}}$ and $b_T=N_{\mathrm{tray}}-\beta_2 b_I$; the maximum-likelihood estimate for $s$ is then simply 
\begin{equation}
\label{eq:s}
s = N_\mathrm{obs} - \beta_1 (N_{\mathrm{tray}}-\beta_2 N_\mathrm{instr}) - \beta_3 N_{\mathrm{instr}}. 
\end{equation}

\begin{figure}[b]
\centering
\includegraphics[width=0.99\textwidth]{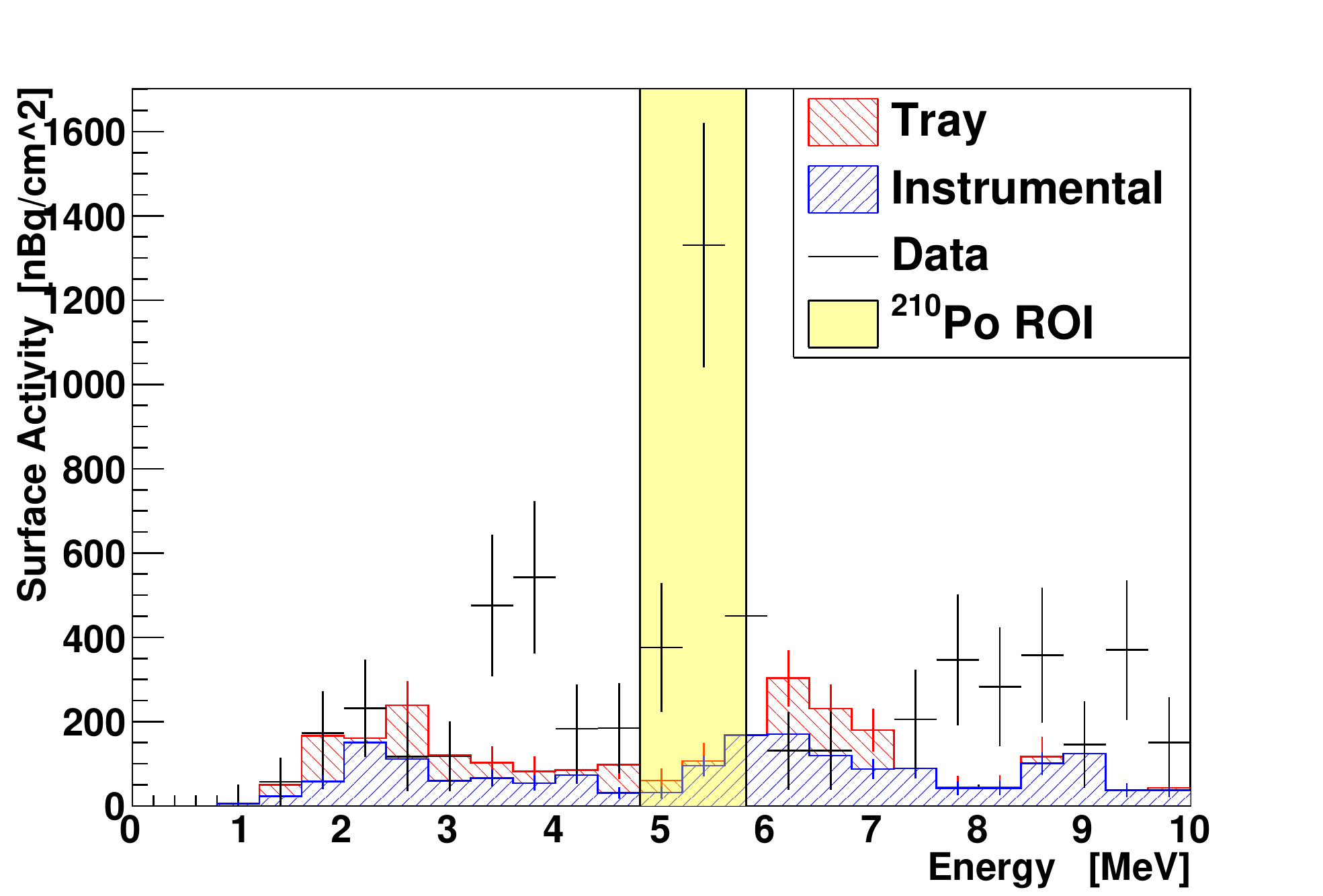}
\vspace{-2em}
\caption{Simulated alpha spectrometer data for a sample that does not cover the full area of the sample tray under the counting anode; thus, there are nonzero contributions from both the instrumental- and tray-background components. The \Po region of interest (ROI) centered at 5.3\,MeV is highlighted.  The background contributions are shown as a stacked histogram, with the tray background (red hatched) stacked on top of the instrumental background (blue hatched) and including uncertainties (red and blue error bars).  Any excess of the simulated data (black error bars) above the total background is attributed to surface activity from the sample.}
\label{fig:XIA_example}
\end{figure}

Figure~\ref{fig:XIA_example} illustrates the contributions from the various components. There is a contribution from the tray background above the instrumental background, and any excess above the total estimated background is attributed to the sample. In general, we measure the tray background in advance of each sample assay for which part of the tray under the counting anode is not fully covered by the sample.  That is, we perform a dedicated measurement of $N_{\mathrm{tray}}$ and the corresponding scaling parameters for each such sample assay. While the conductive Teflon liner  has a relatively low intrinsic alpha activity, as it is used it gradually becomes contaminated with \pb (and thus $^{210}$Po) due to exposure to environmental radon when samples are loaded and unloaded.  This necessitates periodic cleaning or replacement of the Teflon liner. Consequently, the tray background can vary from run to run, thus justifying a dedicated measurement of the tray background prior to each sample assay in which the tray is not fully covered by the sample. On the other hand, the instrumental background is expected to be relatively constant from run to run.  Thus, we model the instrumental background as the average spectral response from several measurements of one of the ultra-high-purity electroformed sheets. For all of the copper sample assays discussed in the remainder of this paper, only the instrumental background is relevant, because $\beta_1$ in Eq.~\ref{eq:s} is equal to zero.

We define the likelihood ratio 
\begin{equation}
\label{eq:ratio}
\lambda(s) = \frac{{\cal L}(s;\hat{\hat{b}}_T,\hat{\hat{b}}_I)}{{\cal L}(\hat{s};\hat{b}_T,\hat{b}_I)},
\end{equation}
where $\hat{\hat{b}}_T$ and $\hat{\hat{b}}_I$ signify the values of the background components that maximize the likelihood for a fixed value of $s$, and ${\cal L}(\hat{s};\hat{b}_T,\hat{b}_{I} )$ is the maximum value of the likelihood function without any constraint on $s$.
With this ratio, we define a test statistic $q(s) = -2\ln{\lambda(s)}$, which is a function of only one parameter. 
According to Wilks' theorem, $q(s)$ will approach a $\chi ^2$ distribution with one degree of freedom \cite{wilks1938}.
We determine the 90\% upper confidence limit on $s$ by scanning $q(s)$ until it reaches 2.71. 
The approximation proposed by Wilks is not entirely suitable for very low event counts, but we still apply it as this sets a weaker upper limit; thus, we are less prone to underestimating a sample's surface activity.

\subsection{Measurement of electroformed copper sheets}\label{sec:PNNL_EF}
\begin{figure}
\centering
\includegraphics[width=0.95\textwidth]{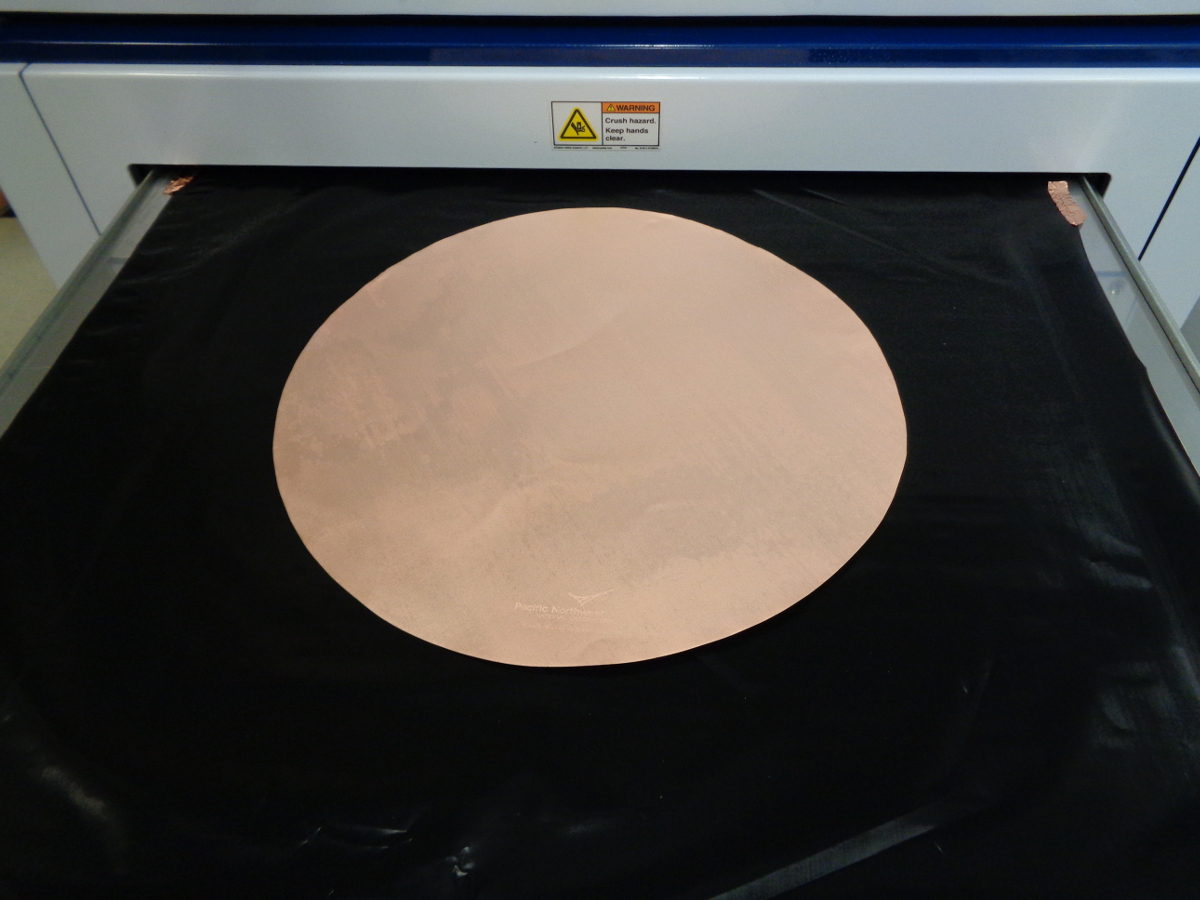}
\caption{Ultra-high-purity PNNL electroformed copper disc ($\sim$30\,cm diameter) placed on the sample tray of the SMU UltraLo-1800 spectrometer prior to a measurement. The conductive (black) Teflon liner is visible underneath and covers the majority of the spectrometer's stainless-steel sample tray.}
\label{fig:PNNL_EF}
\end{figure}

Three different electroformed copper sheets were measured for this work.  The first is a disc with a 30-cm diameter that had been fabricated previously by PNNL and then used by XIA and its customers to evaluate the background performance of their UltraLo-1800 spectrometers (such as the prior alpha-emissivity measurements referred to in Sec.~\ref{sec:cu:ef}). This sheet is shown placed on the SMU sample tray in Fig.~\ref{fig:PNNL_EF}.  For this geometry, the sheet fully covers the part of the tray under the inner electrode; thus, there is no contribution from the tray background when reading out the inner counting anode.  This is the configuration and readout used for most of our measurements. The single exception is the square electroformed sheet, which we additionally measured by reading out the full anode (as discussed below).

\begin{table*}
\begin{tabular}{|c|c|c|c|c|}
\hline
Run start & Measured & Live time & Alphas & Activity in ROI \\
(date) & Side & (hours) & (counts) & (nBq/cm$^2$) \\
\hline
Nov 24, 2015 & Dull & 194.6 & 21.7 & 87.8 $\pm$ 18.8\\ 
Dec 15, 2015 & Dull & 158.7 & 12.5 & 61.7 $\pm$ 17.5\\ 
Dec 22, 2015 & Shiny & 322.8 & 53.6 & 130.4 $\pm$ 17.8\\ 
Jan 12, 2016 & Dull & 149.9 & 10.2 & 53.6 $\pm$ 16.8\\ 
\hline
\multicolumn{5}{|c|}{\textit{After re-etching}}\\
\hline
May 24, 2016 & Dull  & 306.5 & 12.6 & 32.2 $\pm$ 9.1\\ 
Jun 9, 2016 & Shiny & 330.0 & 10.3 & 24.5 $\pm$ 7.6\\ 
Nov 1, 2016 & Dull & 306.4 & 9.1 & 23.5 $\pm$ 7.8\\ 
Jul 21, 2017 & Dull & 54.8 & 4.6 & 65.5 $\pm$ 30.6  \\ 
\hline
Total after re-etching & & 997.7 & 36.6 & 28.8 $\pm$ 4.8\\ 
\hline
\end{tabular}
\caption{Measurement summary for the first electroformed copper disk. The column labeled `Alphas' gives the number of events observed in the \Po ROI (including a small correction for the counting-anode alpha collection efficiency). This number is divided by the live time and sample area to obtain the surface emissivity, which we multiply by 2 to estimate the total surface activity on a given side in the \Po ROI (rightmost column).  Due to exposure to environmental contaminants during prior handling, the surface activity is clearly elevated prior to re-etching of the disc (esp.\ for the `shiny' side). Following application of the acidified-peroxide wet etch, the measurement results are consistently low; the sum (bottom row) is used as the basis for our instrumental-background model.}
\label{tab:PNNL_EF}
\end{table*}

Based on the purity of the electroforming process, our general expectation is that the rate of alphas emitted from the surfaces of the electroformed sheets should be negligibly small.  However, the first disc-shaped sheet was shipped among several institutions prior to our measurements. Such handling means that this particular sheet had a nontrivial exposure to environmental contaminants, particularly on its shinier side which was generally placed facing upward such that it received a larger effective radon and dust exposure.  Measurement results  are summarized in Table~\ref{tab:PNNL_EF}. The elevated levels for the first several measurements are a clear indication of surface contamination, which prompted us to re-etch the disc following the protocols outlined in Sec.\,\ref{sec:cu:clean}.  The four measurements after re-etching gave consistently low results in the \Po ROI, comparable to the first measurements of this disc by XIA~\cite{private_xia} and thus suggesting that the observed rate is representative of the instrumental background.  We combine these four measurements as the basis for our instrumental-background model (i.e., the blue hatched background component in Figs.~\ref{fig:XIA_example} and~\ref{fig:Cu1_spectrum}).  Note that unlike the results for our OFHC copper samples, wet etching of this electroformed disc produced a copper surface without any measurable level of \Po contamination.  Unlike in the OFHC samples, the \Po contamination in the electroformed disc was distributed very near the surface because it was a result of implanted radon progeny. This suggests a different behavior for near-surface versus bulk polonium contamination during the acidified-peroxide etch---a behavior more consistent with the results in Ref.~\cite{HOPPE2007486} for removal of $^{209}$Po \textit{on} copper surfaces.

\begin{table*}[h!]
\begin{tabular}{|c|c|c|c|}
\hline
Run start & Live time & Alphas & Activity in ROI \\
(date) & (hours) & (counts) & (nBq/cm$^2$) \\
\hline
\multicolumn{4}{|c|}{\textit{Second disc-shaped sheet---initial measurements}}\\
\hline
Dec 19, 2017 & 322.9 & 23.9 & 58.3 $\pm$ 11.9\\ 
Mar 26, 2018 & 349.6 & 22.8 & 51.3 $\pm$ 10.7\\ 
Apr 10, 2018 & 664.9 & 28.5 & 33.6 $\pm$ 6.3\\  
\hline
\multicolumn{4}{|c|}{\textit{Second disc-shaped sheet---after re-etching}}\\
\hline
Jan 18, 2019 & 591.8 & 8.0 & 10.6 $\pm$ 3.7\\
Feb 12, 2019 & 678.3 & 20.5 & 23.7 $\pm$ 5.2\\
Mar 13, 2019 & 619.7 & 10.2 & 12.9 $\pm$ 4.1\\
\hline
\multicolumn{4}{|c|}{\textit{Large square sheet---full-anode readout}}\\
\hline
Mar 2, 2018 & 279.5 & 18.7 & 20.7 $\pm$ 4.8\\
Nov 2, 2018 & 448.6 & 47.4 & 32.6 $\pm$ 4.7\\
Nov 21, 2018 & 639.8 & 51.8 & 25.0 $\pm$ 3.5\\
\hline
\multicolumn{4}{|c|}{\textit{Large square sheet---inner-anode readout}}\\
\hline
Dec 18, 2018 & 640.8 & 9.1 & 11.2 $\pm$ 3.7\\
\hline
\end{tabular}
\caption{Measurement summary for the the second electroformed copper disk (30-cm diameter) and the electroformed copper square ($\sim$45$\times$45\,cm$^2$). The columns have the same meanings as in Table~\ref{tab:PNNL_EF}. The disk-shaped sheet was measured by reading out the spectrometer's 707\,cm$^2$ inner counting anode, whereas the square-shaped sheet was measured by reading out either the full 1800\,cm$^2$ counting anode \textit{or} the inner anode. The \Po ROI activities are generally consistent with the instrumental-background model. See main text for further discussion.}
\label{tab:PNNL_EF_liner}
\end{table*}

To construct our instrumental-background model, we assumed that the re-etched electroformed copper is sufficiently pure and clean such that the surface activity is negligible.  One method for testing this assumption is to perform further measurements of electroformed copper to see if they yield consistent results.  
Toward this end, a second set of electroformed sheets was produced, including a square sheet large enough to cover the full tray and a second 30-cm diameter disc (as described in Sec.\,\ref{sec:cu:ef}). 
Measurement results for these two sheets are summarized in Table~\ref{tab:PNNL_EF_liner}. 
The initial surface activity for the second disc was somewhat elevated relative to the levels measured for the first disc (after it was re-etched), suggesting that the initial preparation was not quite as clean.  The first three measurements are consistent (spectrally and temporally) with a low level of \Po contamination; outside the \Po ROI, they are consistent with the instrumental-background model.  This highlights the importance of consistent and careful application of the wet-etching protocols. We re-etched the second electroformed disc and then remeasured it several more times.  Following the re-etch, the results in the \Po ROI are generally consistent with the instrumental-background model. The spectra for a couple of these measurements show a slightly reduced rate at $\sim$5.5\,MeV, which is where we expect any background contributions due radon contamination in the count gas. For this part of the spectrum, we have observed some variation in the background performance that correlates with changeover of the liquid-argon dewar.  This suggests a few-nBq/cm$^2$ systematic uncertainty in the \Po ROI associated with our instrumental-background model.

Finally, we performed measurements of the electroformed copper square.  This sheet is large enough to cover the entire active area of the sample tray, making it possible to evaluate the instrumental background when reading out the full-area counting anode. To compare with the electroformed discs, we also measured the square sheet by reading out the circular inner anode.  The full set of measurements is summarized in Table~\ref{tab:PNNL_EF_liner}. The results are generally consistent with our instrumental-background model.

\subsection{Measurement of McMaster OFHC copper plates}
\label{ssec:mcmaster}

\begin{figure*}[b]
\centering
\includegraphics[width=.49\textwidth]{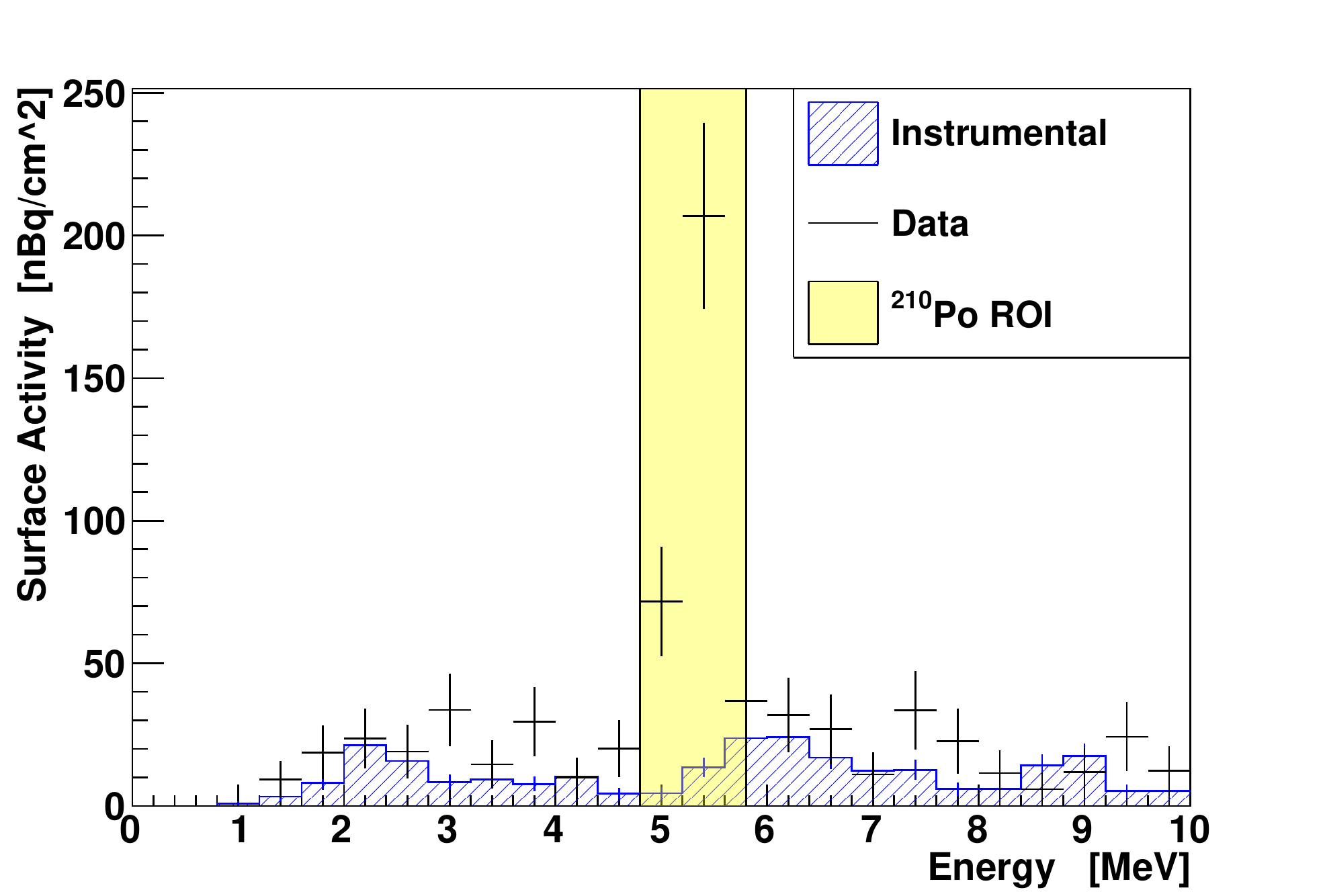}
\includegraphics[width=.49\textwidth]{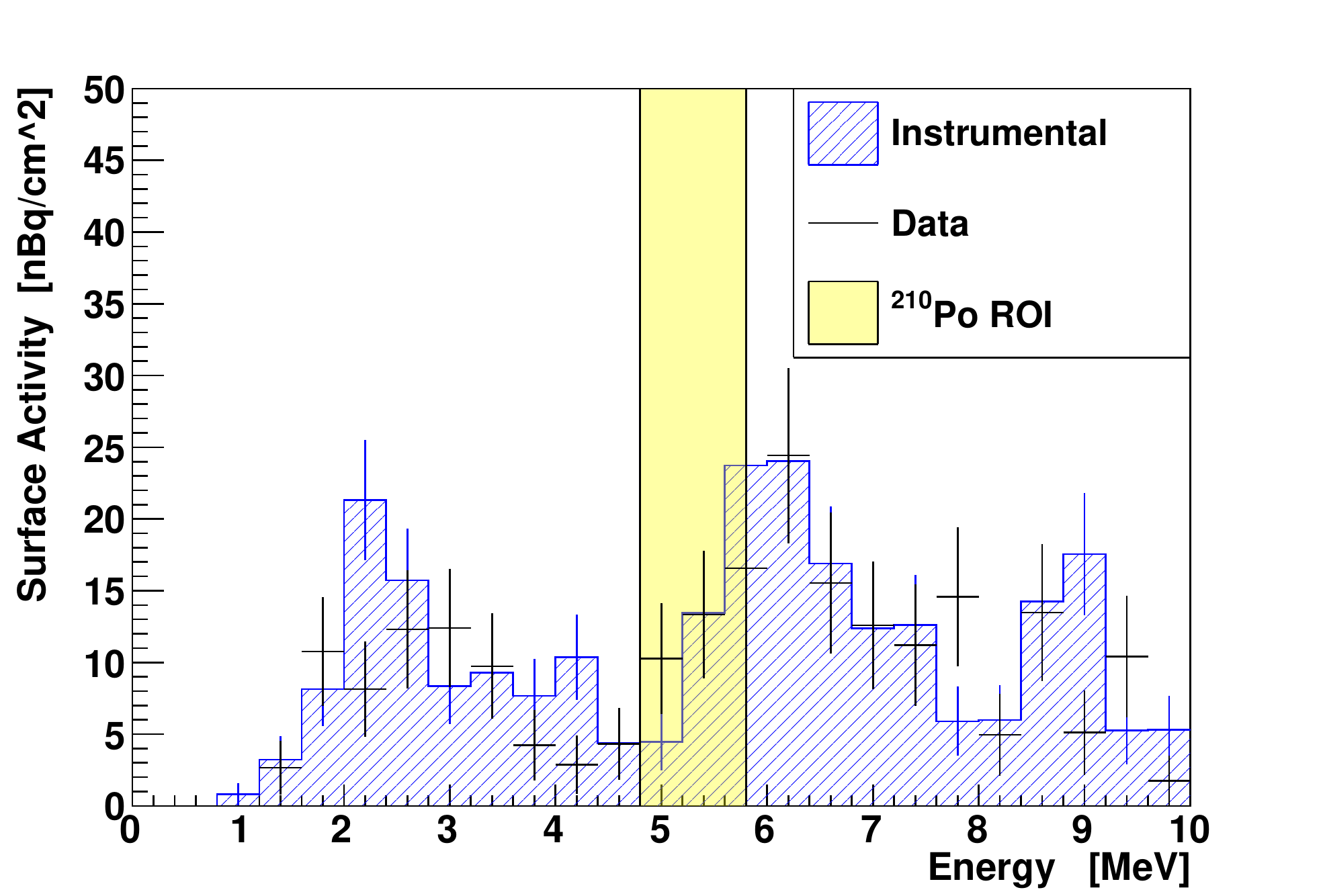}
\caption{Measurement spectra (black error bars) of the McMaster OFHC copper sample compared to the spectrometer's instrumental-background model (blue hatched) and highlighting the \Po ROI. (\textit{left}) First measurement following the initial acidified-peroxide etch (Nov~22, 2016 entry in Table~\ref{tab:Cu1}); there is a clear excess associated with \Po contamination on the surfaces of the copper plates. (\textit{right}) First measurement following application of the electroplating treatment (Sep 6, 2017 entry in Table~\ref{tab:Cu1}); consistency with the background model demonstrates that the electroplating treatment is an effective technique for mitigating \Po surface contamination.}
\label{fig:Cu1_spectrum}
\end{figure*}

Following application of the acidified-peroxide etch, the McMaster OFHC copper plates were measured with the UltraLo-1800 spectrometer three times over a period spanning $\sim$7~months. The results for the \Po ROI are summarized in the first few rows of Table~\ref{tab:Cu1}. The four plates were arranged on the sample tray to form a single square 12~inches to a side and thus fully covered the part of the tray below the inner counting anode.  Consequently, there is no background contribution from the tray; $\beta_1$ is equal to zero in Eq.~\ref{eq:s}. The spectrum for the first measurement is shown in the left panel of Fig.~\ref{fig:Cu1_spectrum}.  There is a clear excess of \Po surface activity above the instrumental background that decreases in the subsequent measurements.

\begin{table*}
\begin{tabular}{|c|c|c|c|c|c|}
\hline
Run start & Live time &  Alphas & Background &  Activity & 90\% Upper Limit\\ 
(date) & (hours) & (counts) & (counts) & (nBq/cm$^2$) & (nBq/cm$^2$)\\
\hline
Nov 22, 2016 & 173.3 & 61.4 & 6.4 $\pm$ 1.1 & 249.6 $\pm$ 35.9 & 312.9\\ 
Mar 10, 2017 & 270.4 & 74.0 & 9.9 $\pm$ 1.6 & 186.1 $\pm$ 25.4 & 232.4\\ 
Jun 13, 2017 & 326.5 & 58.1 & 12.0 $\pm$ 2.0 & 110.9 $\pm$ 18.9 & 144.4\\ 
\hline
\multicolumn{6}{|c|}{\textit{After electroplating treatment}}\\
\hline
Sep 6, 2017 & 604.3 & 22.8 & 22.2 $\pm$ 3.7 & 0.8 $\pm$ 7.8 & 14.3\\ 
Jul 16, 2018 & 328.1 & 16.0 & 12.0 $\pm$ 2.0 & 9.4 $\pm$ 10.7 & 31.1\\ 
\hline
\end{tabular}
\caption{Measurement summary for the McMaster OFHC copper sample. The columns have the same meaning as in Table~\ref{tab:PNNL_EF}. Additionally, the number of events expected in the \Po ROI from the instrumental-background model (`Background') is indicated for each measurement. In this table, the `Activity' is our maximum-likelihood estimate of the sample's \Po surface activity using the formalism outlined in Sec.\,\ref{ssec:subtraction}, and the corresponding 90\% upper confidence limit is provided in the rightmost column. The first three measurements clearly show excess \Po surface activity, which decreases versus time $\sim$ consistent with the \Po half-life.  The final two measurements demonstrate the effectiveness of the electroplating treatment for mitigating this \Po surface activity.}
\label{tab:Cu1}
\end{table*}

\begin{figure}
\centering
\includegraphics[width=\textwidth]{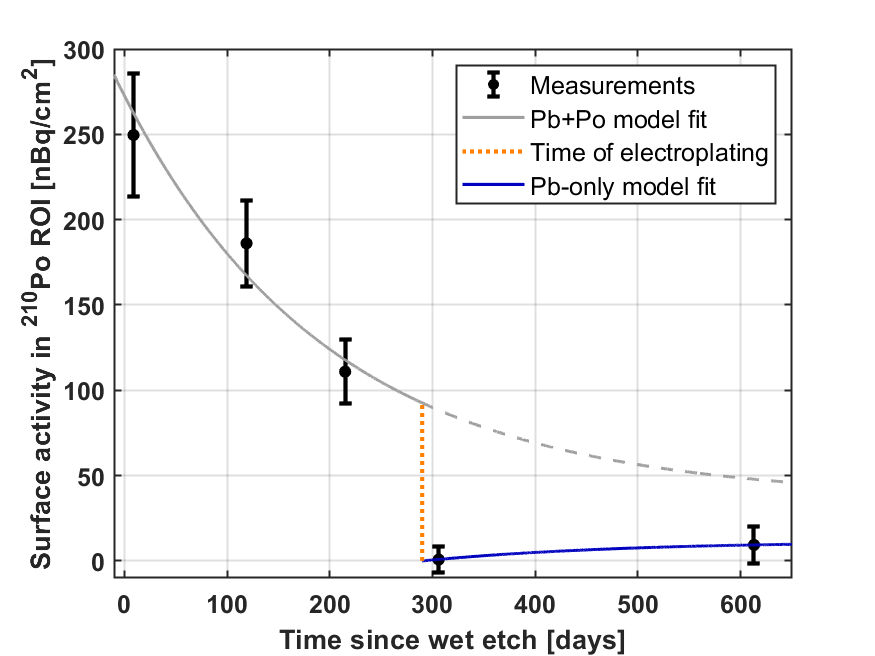}
\caption{Surface activity measurements (error bars) in the \Po ROI for the McMaster OFHC copper sample as a function of time since the initial wet etch. The \Po and \pb surface activities are estimated by fitting the model in Eq.~\ref{eq:Pbfit}  (solid gray), which is also used to estimate the sample's expected \Po surface activity at later times (dashed gray). Following application of the electroplating treatment (see Sec.\,\ref{sec:cu:plate}), the measured activities are consistent with the instrumental background. See main text for further discussion.}
\label{fig:Cu1_assays}
\end{figure}

With three measurements of the \Po activity separated by several months, we can infer the initial levels of \Po and \pb surface contamination immediately after the wet etch. We fit the following model for the sample's \Po surface activity as a function of time:
\begin{equation}
{\cal A}(t) = Ae^{-t/\tau_1} + B(1 - e^{-t/\tau_2}) e^{-t/\tau_3},
\label{eq:Pbfit}
\end{equation}
where $A$ and $B$ are the respective initial levels of \Po and \pb, $\tau_1$  is the \Po mean life (199.7\,days), $\tau_2$ is the sum of the \Po and $^{210}$Bi mean lives (206.9\,days), and $\tau_3$ is the \pb mean life (32\,years)~\cite{SHAMSUZZOHABASUNIA2014561}.  This model fits well to the first three measurements ($\chi_{0}^{2} = 0.8$), yielding best-fit estimates for the initial \Po and \pb surface activities: $A = 272.8\pm35.6$~nBq/cm$^2$ and $B=39.1\pm36.6$~nBq/cm$^2$.  The measurements and the best-fit model are plotted as a function of time in Fig.~\ref{fig:Cu1_assays}.

While the \pb surface contamination is nearly consistent with zero, the initial \Po activity is significant.  Considering the measurements of \pb and \Po in bulk copper in Ref.~\cite{ABE2018157}, we expect that significant amounts of both \pb and \Po are present in the bulk of our OFHC copper plates. Our results suggest that the \pb readily passes from the bulk into solution during the acidified-peroxide etch, whereas the \Po appears to trap at the copper-etchant boundary such that it is effectively concentrated at the copper surface.  
Following this scenario, our best-fit initial \Po surface activity is consistent with being sourced from a \Po concentration in the bulk of the McMaster plates of at least 100\,mBq/kg (at the time of the etch).

Observation of a significant \Po surface activity motivated investigation of an alternative surface treatment based on the PNNL electroforming method.  After the third measurement of the McMaster copper sample, the electroplating treatment described in Sec.\,\ref{sec:cu:plate} was applied to each plate and the full set was measured two more times.  Results from the post-plating measurements are summarized in the bottom rows of Table~\ref{tab:Cu1}; both are consistent with the spectrometer's instrumental background.  Based on the best-fit model (Eq.~\ref{eq:Pbfit}) of the pre-plating measurements, the \Po activity  would have been $\sim$90 and 50\,nBq/cm$^2$ for these two measurements if no treatment had been applied. The spectrum for the first post-plating measurement is provided in the right panel of Fig.~\ref{fig:Cu1_spectrum} and is clearly consistent with the instrumental background.  These results represent a successful demonstration of our electroplating technique for mitigating \Po surface contamination.  As an exercise, we fit a simplified version of Eq.~\ref{eq:Pbfit} to the post-plating results where only the second term was included (i.e., assuming that the post-treatment \Po activity is negligible and thus $A = 0$\,nBq/cm$^2$). The best-fit result is plotted in Fig.~\ref{fig:Cu1_assays} (blue curve) and yields $B=12.2\pm13.8$\,nBq/cm$^2$,  which corresponds to the maximum \pb surface activity that is consistent with the two post-plating measurements.

\subsection{Aurubis copper cleaning and measurement}
\label{ssec:aurubis}

\begin{table*}[b]
\begin{tabular}{|c|c|c|c|c|c|}
\hline
Run start  & Live time &  Alphas & Background &  Activity & 90\% Upper Limit \\
(date) & (hours) & (counts) & (counts) & (nBq/cm$^2$) & (nBq/cm$^2$)\\
\hline
Mar 22, 2017 & 472.9 & 131.0 & 17.3 $\pm$ 2.9 & 188.9 $\pm$ 19.6 & 222.7\\
Jul 24, 2017 & 182.3 & 42.2 & 6.7 $\pm$ 1.1 & 153.2 $\pm$ 28.4 & 206.9\\
Oct 24, 2017 & 502.1 & 56.8 & 18.4 $\pm$ 3.0 & 60.2 $\pm$ 12.7 & 82.9\\
Feb 1, 2018 & 626.8 & 39.8 & 23.0 $\pm$ 3.8 & 21.1 $\pm$ 9.2 & 37.6\\
\hline
\end{tabular}
\caption{Measurement summary for the Aurubis OFHC copper sample. The columns have the same meaning as in Table~\ref{tab:Cu1}. Following the  acidified-peroxide wet etch in late Feb 2017, a significant \Po surface activity is observed that decreases with each successive measurement according to the \Po half-life. These measurements are consistent with the lowest level of \pb contamination ever demonstrated on an OFHC copper surface.}
\label{tab:Cu2}
\end{table*}

Our goal in measuring the Aurubis OFHC copper sample was to characterize the performance of the acidified-peroxide etch for preparing a low-\pb copper surface. Recall that the Aurubis sample was fabricated from the same copper stock---manufacturer and form factor---as will be used to fabricate SuperCDMS SNOLAB detector components (see Sec.\,\ref{sec:cu:vendor}). Additionally, the acidified-peroxide etch will be used as the final surface cleaning treatment for all copper components. Thus, by using the same materials and procedures, our results should be representative of the level of \pb we can expect on the surfaces of SuperCDMS SNOLAB copper detector components.

\begin{figure}
\centering
\includegraphics[width=0.95\textwidth]{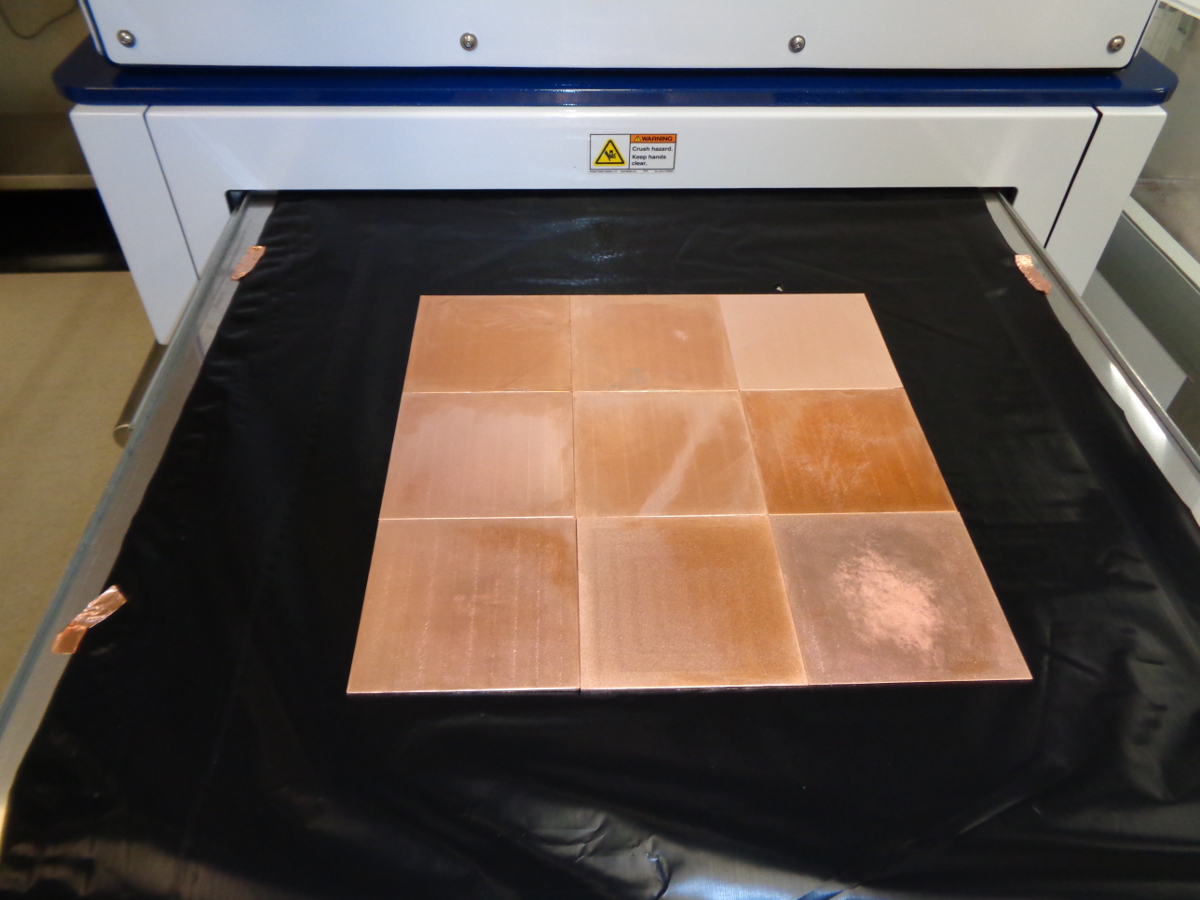}
\caption{The nine Aurubis OFHC copper plates arranged to form a single square (12" to a side) and centered on the spectrometer's sample tray prior to a measurement. The conductive (black) Teflon liner is visible underneath and covers the majority of the spectrometer's stainless-steel tray.}
\label{fig:Cu2}
\end{figure}

\begin{figure}
\centering
\includegraphics[width=\textwidth]{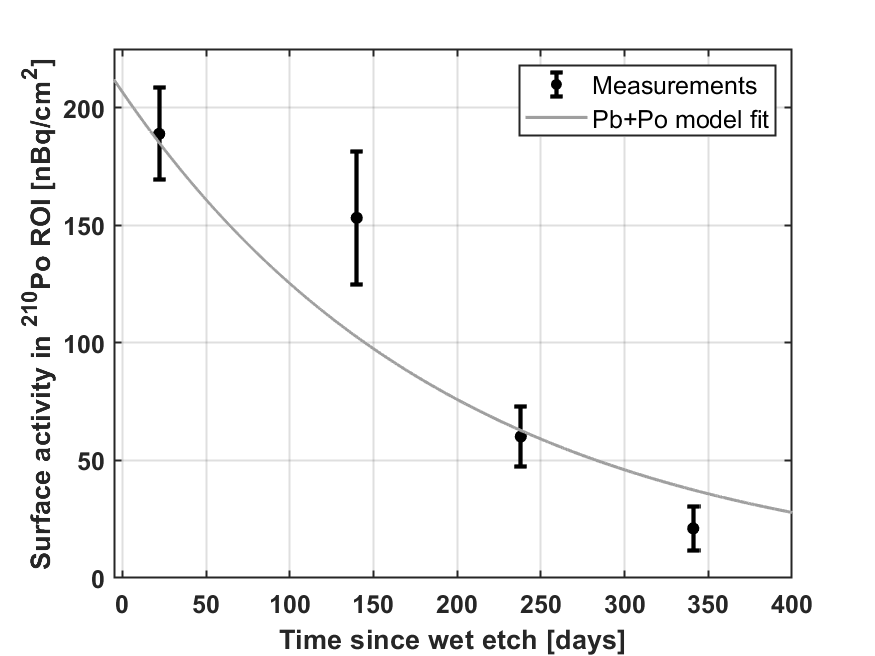}
\caption{Surface activity measurements (error bars) in the \Po ROI for the Aurubis copper sample versus time since the wet etch.  Fitting the model in Eq.~\ref{eq:Pbfit} (solid gray) yields estimates of the initial \Po and \pb surface activities. The former is significant while the latter is consistent with zero.}
\label{fig:Cu2_assays}
\end{figure}

The Aurubis OFHC copper plates are shown in Fig.~\ref{fig:Cu2}, centered on the spectrometer's sample tray and arranged so as to fully cover the part of the tray below the inner counting anode.  Thus, similar to measurement of the McMaster sample, there is no background contribution from the tray when measuring the Aurubis plates; $\beta_1$ in Eq.~\ref{eq:s} is equal to zero. Following the acidified-peroxide etch, we measured the Aurubis sample four times over a period spanning nearly a full year; results for the \Po ROI are summarized in Table~\ref{tab:Cu2}.  Having four measurements separated by such large intervals of time enables a more precise determination of the levels of \Po and \pb contamination following the wet etch. We again use the model in Eq.~\ref{eq:Pbfit} to fit to the \Po activity as a function of time, yielding best-fit estimates for the initial \Po and \pb surface activities:  $A=206.7\pm21.4$\,nBq/cm$^2$ and $B=0\pm12.2$\,nBq/cm$^2$, where the activities were constrained to be positive. To our knowledge, this is the lowest level of \pb ever demonstrated on an OFHC copper surface. Similar to the McMaster sample, we observe a significant initial \Po surface activity, suggesting that the Aurubis copper also has an appreciable concentration of \Po in the bulk material.
Note that the best-fit model is not perfectly consistent with the measurements ($\chi^{2}_{0} = 3.2$); both are plotted in Fig.~\ref{fig:Cu2_assays}. Based on the measured spectra, this appears to be due to upward fluctuations of the spectrometer's instrumental background as a result of elevated radon levels in the count gas for the first and second measurements.

\section{Bulk $^{210}$Pb assay of copper}\label{sec:bulk210Pb}

The results of the prior section strongly suggest the presence of \Po \emph{in the bulk} of the McMaster and Aurubis copper samples, which appears consistent with the bulk levels reported for OFHC copper in Ref.~\cite{ABE2018157}. Ref.~\cite{ABE2018157} further reports that OFHC copper can contain $^{210}$Pb in the bulk at levels from 17--40~mBq/kg. It is natural to expect that our OFHC copper samples might have a similar level of \pb in the bulk. 

The measurements in Ref.~\cite{ABE2018157} also used an UltraLo-1800 spectrometer.  However, a much different method was used to prepare the surfaces of their copper samples~\cite{ABE2018157,ZUZEL2012140} compared to the surface treatment methods reported here. While our measurements focus on \Po alphas emitted from the \textit{surface} of the etched copper samples, their measurements focus on a lower-energy ROI in the alpha spectrum that is sensitive to \Po alphas emitted from the \textit{bulk} of the copper samples. The smooth finish achieved by their surface preparation allows for a well-calibrated measurement of  the bulk \Po concentration based on this lower-energy ROI.  While it is (in principle) possible to analyze our UltraLo-1800 data in a similar way, our surface preparation methods tend to yield rougher surfaces and thus the surface-to-bulk calibration factor determined in Ref.~\cite{ABE2018157} is not directly applicable.

We have pursued an alternative approach to assay for \pb in bulk copper. In $\sim$4\% of \pb decays, a 46.5\,keV gamma-ray is emitted~\cite{SHAMSUZZOHABASUNIA2014561}. This gamma-ray can be used to directly quantify the \pb concentration in a copper sample.  In this section, we describe our efforts to perform such measurements in order to directly confirm the bulk levels of \pb in our OFHC copper samples.

\subsection{Bulk $^{210}$Pb assay method}

\begin{table}[t]
\begin{tabular}{|c|c|c|c|}
\hline
       & Square & Plate & Total \\
Sample & Dimension & Thickness & Mass \\
       & (mm) & (mm) & (g) \\
\hline
McMaster & 152.4 & 5 & 1250 \\
Aurubis & 101.6 & 2 ($\times3$) & 568 \\
\hline
\end{tabular}
\caption{OFHC copper samples $\gamma$ counted for the presence of $^{210}$Pb.}
\label{tab:CuCounted}
\end{table}

A subset of the McMaster copper plates and a subset of the Aurubis copper plates were separately counted on the Roseberry 6530 S-ULB germanium detector~\cite{6530inPrep} at the Boulby Underground Laboratory~\cite{SCOVELL2018160}. This detector is a Mirion Technologies (formerly Canberra) BEGe-style BE6530 high purity germanium (HPGe) detector, ideally suited for $\gamma$- and x-ray measurements in the 20--100~keV energy range. The 46.5~keV \pb $\gamma$-ray is readily attenuated by the thickness of the copper sample itself.  
A simple \textsc{Geant}4~\cite{1610988,AGOSTINELLI2003250} simulation was used to estimate that the closest $\sim$0.75~mm thickness of a copper sample placed on top of the detector will contribute 95\% of the $\gamma$-rays measured in the 46.5~keV peak.  
This suggests that a copper sample thickness of only a few mm provides more than sufficient counting mass for measurement of the \pb 46.5~keV $\gamma$-rays. Consequently, just one of the relatively thick McMaster plates was chosen for measurement, whereas three of the Aurubis plates were stacked for measurement. 
The details of the counted copper samples are provided in Table~\ref{tab:CuCounted}.  Note that prior to measurement of the McMaster copper plate, the outer layer of electroplated copper (see Sec.\,\ref{sec:cu:plate}) was milled off in order to re-expose the original, underlying OFHC copper material.

\subsection{$^{210}$Pb calibration with lead standards}
To determine the \pb analysis ROI for the Roseberry detector, lead samples containing previously (and independently) measured concentrations of \pb were considered prior to measurement of the copper samples. These lead `standards' span a range of concentrations from 0.6--330~Bq of \pb per kg of lead~\cite{Orrell2016,KEILLOR2017185}. Each \pb standard is a solid, rectilinear piece of lead measuring 1$\times$10$\times$10~cm$^3$.  In particular, two of the highest-activity standards were selected to calibrate the HPGe detector:  75 and 330~Bq/kg.  Results from the two calibration measurements of the Roseberry detector's response to 46.5~keV \pb $\gamma$-rays are presented in Fig.~\ref{fig:Pb210Standards}.

\subsection{Bulk $^{210}$Pb assay results}
From November 2018 to January 2019, the copper samples were counted and a background spectrum was collected. The McMaster plate was counted for 13.2~days and the Aurubis plates were counted for 23.6~days. The detector's  background was measured for a period of 23.4~days. 
Analysis of the collected data included use of a \textsc{Geant}4 simulation of the Roseberry detector's detection efficiency for 46.5~keV $\gamma$-rays emitted from $^{210}$Pb decays in both the copper samples and the lead calibration sources, matching the actual sample geometries.

\begin{figure}
\centering
\includegraphics[width=\columnwidth]{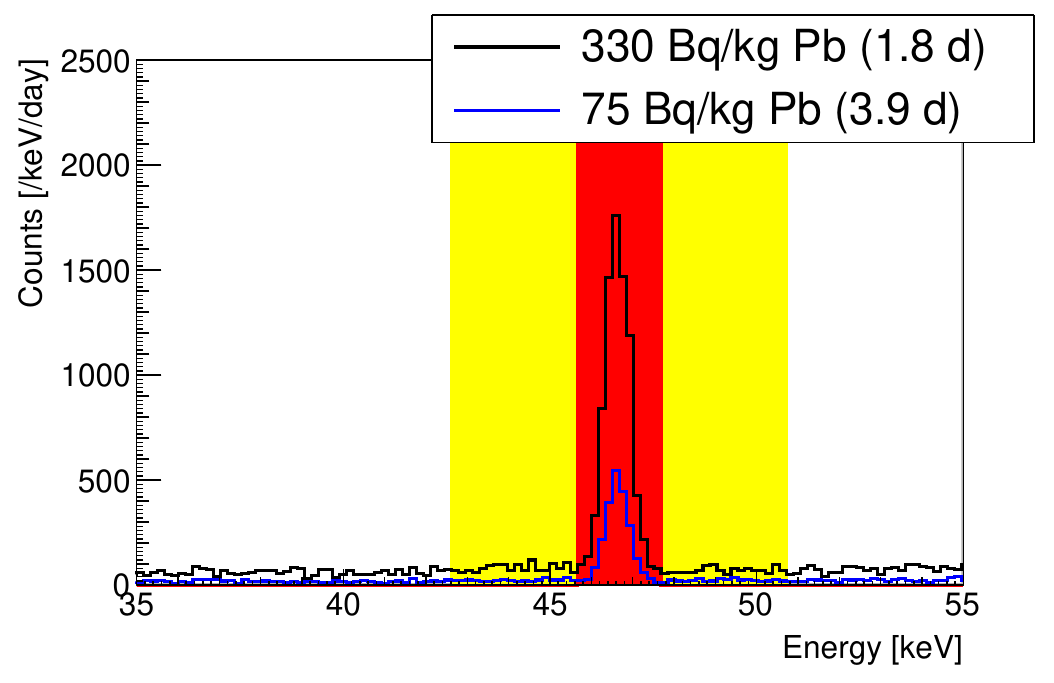}
\caption{Roseberry HPGe detector spectra measured for the 330 and 75~Bq/kg \pb lead standards. The area selected to represent the background is highlighted in yellow (left and right of the peaks) and the peak area is highlighted in red. \textit{Figure courtesy of Paul R.\ Scovell and Emma Meehan, Boulby Underground Laboratory.}} 
\label{fig:Pb210Standards}
\end{figure}

Using the background and peak regions indicated in Fig.~\ref{fig:Pb210Standards}, the measurement of the McMaster copper sample yielded only an upper limit: $<17$~mBq of $^{210}$Pb per kg of copper. In the case of the Aurubis copper sample, the analysis resulted in a measurement of $21\pm15$~mBq of $^{210}$Pb per kg of copper. From these results we generally infer that the $^{210}$Pb concentrations in our two  samples are at or below the Roseberry detector's sensitivity level for measuring \pb in bulk copper, corresponding to $\sim$20~mBq of $^{210}$Pb per kg of copper.

\subsection{Bulk contaminants discussion}
Knowing the bulk \pb concentrations is potentially useful for correlating to the post-cleaning $^{210}$Po surface activities (reported in Sec.\,\ref{sec:xia}) that resulted from application of the acidified-peroxide wet etch to our copper samples. 
Unfortunately, our goal of measuring the bulk \pb via gamma counting was not achieved.  Despite the excellent sensitivity of the Roseberry detector, the results are essentially consistent with setting an upper limit at the instrument's sensitivity level. However, because the detector's sensitivity to \pb in bulk copper is at the level of a few tens of mBq/kg, we also conclude that the bulk \pb levels in our samples are in line with the bulk assay results for OFHC copper in Ref.~\cite{ABE2018157}.

Regarding the relatively high post-etching \Po activities observed on the surfaces of our samples, our hypothesis is that bulk polonium contaminants do not readily pass into the etching solution. During the cleaning process, lead and polonium are liberated from the copper bulk by the acidified peroxide.  
This procedure creates a diffusion layer (also called a Nernst layer) at the surface~\cite{echem_book} as the copper (and lead) are oxidized into the etching solution by the free radicals in the acidified peroxide. It is tempting to conclude that the polonium also goes into solution and then simply redeposits on the copper surface, because it possesses the most reductive potential of these species: E$^{\circ}$ of Po $= +0.76$\,V, Cu $= +0.34$\,V, and Pb $= -0.13$\,V. 
However, our measurements of the electroformed copper sheets demonstrate that \Po contamination can be effectively removed with the acidified peroxide when the contamination is located on or very near the copper surface.  This was also shown previously in Ref.~\cite{HOPPE2007486}, for removal of $^{209}$Po that had been placed directly on copper surfaces. 
Conversely, when polonium is present in the copper's bulk (such as with our OFHC samples), our measurements indicate that it is difficult to remove and can even accumulate at the copper surface. We suspect that a large fraction of the polonium located within the bulk may simply remain in the reduced form as copper is removed around it. If the bulk polonium is oxidized during the etching process, it may be readily reduced near the copper surface where it encounters the electron-enriched diffusion layer that is present during the wet etch. Further, it is known that this cleaning process is most effective when the copper samples are mechanically agitated in the etching solution; the hydrodynamics serve to reduce the thickness of the diffusion layer, thereby increasing transfer of contaminants into the etching solution.
We therefore conclude that the surface cleaning effectiveness depends not only on the bulk $^{210}$Pb and \Po concentrations, but also on the etching solution strength, dwell time in the etching bath, level of agitation of the etching solution, and even the geometries of the copper sample and etching bath. While some of the mechanistic details remain as open questions, the implications of the work presented in this paper nonetheless assist in demonstrating the production of a clean, low-background OFHC copper surface for use in the SuperCDMS SNOLAB experiment.

\section{Conclusions}

We have explored preparation of two OFHC copper samples using two surface treatment methodologies:  an acidified-peroxide wet etch and copper electroplating based on PNNL's electroforming technique.  Surface cleanliness with respect to \pb and \Po was evaluated using a high-sensitivity assay method, developed and performed with an UltraLo-1800 alpha spectrometer.  Although significant \Po levels were measured on the surfaces of both samples following the wet etch, the measured \Po rates versus time are consistent with near-zero levels of \pb.  In particular, for the Aurubis copper sample, our measurements represent the lowest level of \pb ever demonstrated on an OFHC copper surface. These results also demonstrate that the SuperCDMS SNOLAB background goal for \pb contamination on the surfaces of the detector copper housings~\cite{sensitivity} can be achieved using our acidified-peroxide wet etching protocols applied to commercially sourced OFHC copper.

We attribute the post-etching surface alpha activities to \Po contamination present in the bulk of the OFHC copper samples. As the copper is etched, both lead and polonium are released from the bulk material.  While the lead tends to move into the etching solution, the polonium is unable to pass through the diffusion layer at the copper-etchant boundary. If the polonium is oxidized during the etch, it is then efficiently reduced due to its high reduction potential and the availability of copper-oxidation electrons at this layer. On the other hand, polonium located on or near the copper surface is much more susceptible to oxidation and thus removal into the etching solution is observed. We have demonstrated that \Po contamination on the surface of electroformed copper can be mitigated using agitation with the wet etching process. Our conclusions are consistent with the \pb and \Po assay results reported in Ref.~\cite{ABE2018157} for bulk OFHC copper.  Though less sensitive, our attempts to directly measure bulk \pb in our copper samples are similarly suggestive.  

We also explored use of the PNNL copper electroforming technique as a surface treatment method.  Thin layers of ultra-high-purity copper were plated onto lower-purity OFHC copper plates.  Results from subsequent surface assays indicate near-perfect mitigation of the \Po contamination that was present on the copper surfaces prior to the electroplating treatment. This electroplating treatment represents a new method for mitigating \pb \textit{and} \Po on the surfaces of copper components fabricated from commercially sourced copper; an even lower-background copper surface can be achieved than is possible by simply etching OFHC copper. The additional electroplated copper thickness acts as an ultra-high-purity shielding layer to protect detectors from radiocontaminants on the surface \textit{and} in the bulk of the underlying, lower-purity copper. We anticipate that this new method will be useful for controlling copper-related radiogenic backgrounds in future rare-event searches.

\section*{Acknowledgement}
The authors thank Chamkaur Ghag at the University College London and Paul R.\ Scovell and Emma Meehan (Boulby Underground Laboratory) for assisting in this work by performing the measurement of bulk $^{210}$Pb in the copper samples using the Roseberry detector (see Sec.\,\ref{sec:bulk210Pb}). 
We gratefully acknowledge support from the U.S.\ Department of Energy (DOE) Office of High Energy Physics and from the National Science Foundation (NSF). This work was supported in part under NSF Grant No.\ 1707704. Pacific Northwest National Laboratory (PNNL) is operated by Battelle Memorial Institute for the DOE under Contract No.\ DE-AC05-76RL01830. SLAC is operated under Contract No.\ DEAC02-76SF00515 with the United States Department of Energy.

\section*{References}
\bibliography{references}

\end{document}